\documentclass[aps,preprint,nofootinbib,floatfix]{revtex4-1}
\usepackage{graphicx,color,hyperref}
\usepackage{amsmath,amssymb}
\usepackage{url}
\usepackage{epstopdf}
\newcommand{\lsim}{\mathrel{\mathop{\kern 0pt \rlap
  {\raise.2ex\hbox{$<$}}}
  \lower.9ex\hbox{\kern-.190em $\sim$}}}
\newcommand{\gsim}{\mathrel{\mathop{\kern 0pt \rlap
  {\raise.2ex\hbox{$>$}}}
  \lower.9ex\hbox{\kern-.190em $\sim$}}}

\begin{document}

\title{Collider Constraints on the Dark Matter Interpretation\\ 
of the CDMS II Results}

\author{Kingman Cheung$^{1,2}$, Chih-Ting Lu$^2$, Po-Yan Tseng$^{2}$, 
  and Tzu-Chiang Yuan$^3$}

\affiliation{
$^1$Division of Quantum Phases \& Devices, School of Physics, 
Konkuk University, Seoul 143-701, Republic of Korea \\
$^2$Department of Physics, National Tsing Hua University, 
Hsinchu 300, Taiwan\\
$^3$Institute of Physics, Academia Sinica, Nankang, Taipei 11529, Taiwan
}

\date{\today}

\begin{abstract}
  The recent observation of three events by the CDMS II experiment can
  be interpreted as a 8.6 GeV dark matter scatters elastically with the
  nucleons inside the silicon detectors with a spin-independent cross section of
  $1.9 \times 10^{-41}$ cm$^{2}$.  We employ the effective dark matter
  interaction approach to fit to the interpreted cross section, and
  make predictions for monojet and monophoton production at the LHC
  with the fitted parameters.  We show that some of the operators are
  already ruled out by current data while the others can be further
  probed in the upcoming 14 TeV run of the LHC.
\end{abstract}

\maketitle


\section{Introduction}

Evidences for the omnipresence of dark matter in our present Universe
have been well established through its gravitational effects spread
over many different scales, ranging from the rotation curves of clusters
and spiral galaxies, bullet clusters, weak lensing effects to
the cosmic microwave background radiation.  A number of recent
observational experiments, especially the very precise measurement of
the cosmic microwave background radiation in the Wilkinson Microwave
Anisotropy Probe (WMAP) \cite{wmap} and Planck Mission experiments
\cite{planck} suggest that the dark matter (DM) relic density
$\Omega_{c} h^2 = 0.1199 \pm 0.0027$, while the baryon density
$\Omega_b h^2 = 0.02205 \pm 0.00028$ \cite{planck}.  However, its
particle nature remains alluring to theorists and many particle dark
matter models have been built over the years.  In most of these
popular models, dark matter is assumed to be nonbaryonic and
electrically neutral.  Detailed predictions of these models can now be
further scrutinized by using data from various direct and indirect
detection experiments as well as collider experiments like the Large
Hadron Collider (LHC), and useful constraints can be deduced.

Many experiments are being performed or will be carried out in
foreseeable future to investigate the particle nature of the DM. One category
is the direct detection experiments with a large detector buried
deeply underground or underneath
a high mountain. In such background-free
environments, one hopes to detect the rarely happened scattering of
the DM particle with nuclei of the detector materials. A number of
recent experiments detect some events, which cannot be accounted for
by any known background sources, thus are interpreted as signals of DM.
Coincidentally or not, they all fall into the light DM mass region,
around 10 GeV.  The first claimed DM signal was the seasonal variation
in detection rate recorded by the DAMA \cite{dama} experiment. 
Some positive results were also reported 
by CoGeNT \cite{cogent} and CRESST \cite{cresst}. More recently
the CDMS II \cite{cdms} collaboration has seen three events, which
correspond to a DM mass of 8.6 GeV and a spin-independent cross
section of $\sigma^{SI}_N = 1.9 \times 10^{-41} \, {\rm cm}^2$. Such a
large cross section of the DM particle with a nucleon could imply a
large production cross section of DM particles at the LHC. The main
purpose of this note is to investigate were the CDMS II events
interpreted as signals of the DM, what would be the detectable
signature at the LHC?

We adopt an effective interaction approach 
\cite{global,Kurylov:2003ra,cao,lhc-3,lhc-5,lhc-4,ding-liao} 
to describe the interactions of the dark matter particle with the
standard model (SM) particles.  The DM scattering off a nucleon goes 
through the process $q \chi \to q \chi$ while one can interchange
the quark in the final state with the $\chi$ in the initial and 
the process becomes $q\bar q \to \chi \bar{\chi}$. 
Since the DM particles in the final state would escape from detection in the
particle
detectors at the LHC, one has to attach either a gluon, photon, $Z$, or
a $W$ boson in order to give a detectable signal.
Thus, the most anticipated signals of dark matter at hadronic
colliders are large missing energies in association with jets, photons,
or leptons (from $W$ or $Z$ decays). 
For example, if we take one of the operators, $(\bar \chi
\chi)(\bar q q)$, and attach a gluon or a photon to a quark leg, it
will give rise to a monojet or a monophoton plus missing energy
event. The LHC experiments have been
actively searching for these signatures in some other context, such as
large extra dimensions.  We will use the most
updated data on monojet and monophoton production from the LHC
\cite{atlas,cms} to constrain the effective dark matter interactions.

We will consider various spin nature of the dark matter
particle including Dirac and Majorana for fermionic dark matter,
as well as real and complex scalar. Our strategy is as follows.  
For each operator that can contribute to the spin-independent (SI) cross
section between the DM particle and the nucleon, we calculate the
size of the effective scale $\Lambda$ that can account for the 
CDMS II cross section of
$1.9\times 10^{-41} \, {\rm cm}^2$ with dark matter mass of
8.6 GeV.
With these parameter values we calculate
the monojet and monophoton cross sections at the LHC and compare with
the existing data. We shall then repeat the exercise for the operators
that can contribute to the spin-dependent (SD) cross section if the CDMS II
data is to be interpreted due to spin-dependent scattering between
the DM particle and the nuclei \cite{dep}.

The organization of this paper is as follows. In the next section, we 
describe the effective interaction approach and the operators that give rise
to spin-independent and spin-dependent scattering between the DM particle
and the nucleon. In Sec. III, we give the formulas for the DM-nucleon 
scattering cross sections.
In Sec. IV, we determine the best fitted value of the effective scale
of each operator from the CDMS II data and use these best fitted parameters
to calculate the monojet/monophoton cross sections at the LHC. 
We summarize in Sec. V.


\section{Effective Dark Matter Interactions}

We assume that the dark matter particle, denoted by $\chi$, is a 
standard model singlet, and the $\chi$ can stand for a Dirac or 
Majorana fermion,  real or complex scalar, depending
on the context. Also, $f$ stands for a SM fermion, including quarks and
leptons. A thorough discussion of the operators can be found in 
Ref.~\cite{global}. Here we highlight those operators which are relevant to
the spin-independent and spin-dependent scattering between the DM particle
and the nucleon.

In the notation of Ref.~\cite{global}, the operators that contribute
to spin-independent cross sections are:
\begin{eqnarray}
O^{D}_1 & = & \sum_f \frac{C_1^f}{\Lambda_1^2} 
 \left( \bar \chi \gamma^\mu \chi \right)  
  \left( \bar f \gamma_\mu f \right) \; ,\\
O^{D,M}_7 & = & \sum_f \frac{C_7^f m_f}{\Lambda_7^3} 
 \left( \bar \chi \chi  \right) \left( \bar f f \right) \; ,\\
O^{D,M}_{11} & = &\frac{C_{11}}{\Lambda_{11}^3} 
\left( \bar \chi \chi  \right)
\left( - \frac{\alpha_s}{12 \pi} G^{\mu\nu} G_{\mu \nu} \right) \; ,\\
O^{C}_{15} & = & \sum_f \frac{i C_{15}^f}{\Lambda_{15}^2} 
\left( \chi ^\dagger \overleftrightarrow {\partial_\mu} \chi  \right) 
  \left( \bar f \gamma^\mu f  \right) \; ,\\
O^{C,R}_{17} & = & \sum_f \frac{C_{17}^f m_f}{\Lambda_{17}^2} 
 \left( \chi^\dagger \chi  \right) \left( \bar f f  \right) \; ,\\
O^{C,R}_{19} & = &\frac{C_{19}}{\Lambda_{19}^2} \left( \chi^\dagger \chi  \right)
\left( - \frac{\alpha_s}{12 \pi} G^{\mu\nu} G_{\mu \nu} \right) \; ,
\end{eqnarray}
where $\Lambda_i$ is the heavy mass scale for the connector sector 
that has been  integrated out and $C_i$ is an effective dimensionless 
coupling constant 
of order $O(1)$ that can be absorbed into $\Lambda_i$.
Here $D$ and/or $M$ in the superscript of $O_1$, $O_7$, and $O_{11}$ denote
that the DM $\chi$ can be a Dirac and/or Majorana fermion. 
Also, $C$ and/or $R$ in the superscript of $O_{15}$, $O_{17}$, and $O_{19}$ 
denote that the DM $\chi$ can be a complex and/or real scalar.
The $m_f$ dependence in the coupling strength of some of the operators 
is included for  scalar-type interactions because this factor appears 
naturally from  dark matter models with scalar exchange diagrams.
For operators involving gluons, the factor of strong coupling constant 
$\alpha_s(2 m_\chi)$ 
is also included because these operators are induced at one loop level as 
a result of 
integrating the heavy quarks and evaluated at the scale $2 m_\chi$ where 
$m_\chi$ 
is the dark matter mass.

On the other hand, the following operators contribute to the spin-dependent 
scattering cross section
\begin{eqnarray}
O^{D,M}_4 & = & \sum_f \frac{C_4^f}{\Lambda_4^2} 
  \left( \bar \chi \gamma^\mu \gamma^5 \chi \right) 
 \left( \bar f \gamma_\mu \gamma^5 f \right) \; ,\\
O^{D}_5 & = & \sum_f \frac{C_5^f}{\Lambda_5^2} 
  \left( \bar \chi \sigma^{\mu \nu} \chi \right) 
  \left( \bar f \sigma_{\mu\nu} f \right) \; .
\end{eqnarray}

The relative importance to SI or SD scattering cross section from 
each of the above operators can be
easily understood by nonrelativistic expansions, which had been fully analyzed
in previous work \cite{global}. We will ignore the evolution effects 
of the above effective operators in our analysis. 

The validity and pitfalls of using effective dark matter interaction
approach for LHC studies have been examined by a number of authors in
Refs.\cite{eftvalidity-1,eftvalidity-2}.


\section{Direct Detection}

The solar system moves around in the Galactic halo with a
nonrelativistic velocity $v \sim 10^{-3} c$.  
When the dark matter particles move through a detector,
which is usually put under a deep mine or a mountain to reduce
backgrounds, and create collisions with the detector, 
some signals may arise in phonon-type, scintillation-type, 
ionization-type, or some combinations of them, depending on the 
detector materials.
The event rate is extremely low because of the
weak-interaction nature of the dark matter. There are controversies
among various direct detection experiments.  Both 
DAMA \cite{dama} and CoGeNT \cite{cogent} observed some positive 
signals of dark matter detection, 
which
point to a light dark matter ($\sim 5-10$ GeV) with the 
$\sigma^{\rm SI}_{\chi N} \sim
10^{-41} \;{\rm cm}^2$.  The very recent CDMS \cite{cdms} 
has seen three events, which correspond to a DM mass of 8.6 GeV and a 
spin-independent cross section of 
$\sigma^{\rm SI}_{\chi N} = 1.9 \times 10^{-41} \, {\rm cm}^2$
between the DM particle and the nucleon
or a spin-dependent cross section of 
$\sigma^{\rm SD}_{\chi n} = 10^{-35} \, {\rm cm}^2$ \cite{dep}
between the DM particle and the neutron.
We shall use these cross sections and interpret it as 
a SI or SD scattering between the DM particle and the nucleon, and 
calculate the parameter of each operator that can give such cross sections.

In the following we will not concern about the exclusions by 
the XENON100 data \cite{xenon} for spin-independent cross 
sections ($\sigma^{\rm SI}$), and 
XENON10 \cite{xenon-sd}, ZEPLIN \cite{zeplin} and SIMPLE \cite{simple} data
for spin-dependent cross sections ($\sigma^{\rm SD}$). As pointed out
by a few recent analyses that there may be some inconsistency in the low
DM mass region of the XENON data \cite{unbearable}.


\subsection{Spin-Independent Cross Section}

For a nuclei $\mathcal N$ composed of $Z$
protons and $(A - Z)$ neutrons, the SI cross section contributed 
by the operator $O^D_1$ is given by
\begin{equation}
\label{sixs1dirac}
\sigma^{\rm SI}_{\chi \mathcal N} (0) = 
   \frac{\mu^2_{\chi \mathcal N}}{\pi} \vert b_{\mathcal N} \vert^2 \; ,
\end{equation}
where 
\begin{equation}
\mu_{\chi \mathcal N} = \frac{m_\chi m_{\mathcal N}}{m_\chi + m_{\mathcal N}} \; ,
\end{equation}
is the reduced mass for the $\chi {\cal N}$ system and
\begin{equation}
b_{\mathcal N} = Z \, b_p + (A - Z) \, b_n \; ,
\end{equation} 
with
\begin{eqnarray}
\label{sicouplings1a}
b_p & = & \frac{1}{\Lambda_1^2} \left( 2 \, C_1^u + C_1^d  \right) \; ,\\
\label{sicouplings1b}
b_n & = & \frac{1}{\Lambda_1^2} \left( C_1^u + 2 \, C_1^d  \right) \; .
\end{eqnarray}

For $O^D_7$, we have
\begin{equation}
\label{sixs7dirac}
\sigma^{\rm SI}_{\chi \mathcal N} (0) = \frac{\mu^2_{\chi \mathcal N}}{\pi} \vert f_{\mathcal N} \vert^2 \; ,
\end{equation}
where 
\begin{equation}
\label{fN}
f_{\mathcal N} = Z \, f_p + (A - Z) \, f_n \; ,
\end{equation} 
with
\begin{equation}
\label{sicouplings7}
f_{p,n} = \frac{m_{p,n}}{\Lambda_7^3} 
\left\{
\sum_{q = u,d,s} C_7^q \, f^{(p,n)}_{Tq} + \frac{2}{27} f^{(p,n)}_{TG} \sum_{Q=c,b,t} C_7^Q 
\right\} \; ,
\end{equation}
and
\begin{equation}
f^{(p,n)}_{TG} \equiv 1 - \sum_{q=u,d,s} f^{(p,n)}_{Tq} \; .
\end{equation}
For the Majorana case of $O^M_7$, one should multiply 
Eq.(\ref{sixs7dirac}) by a factor of 4.
For a recent re-evaluation of the hadronic matrix elements
$f^{(p,n)}_{Tq}$ using the 
latest lattice calculation results of the strange quark $\sigma_s$ term
and its contribution inside the nucleon, see Ref.\cite{Cheng-Chiang}.

For  $O^D_{11}$, the cross section is the same as $O^D_7$ with the
following couplings
\begin{equation}
\label{fpnO11}
f_{p,n} = \frac{m_{p,n}}{\Lambda_{11}^3} \frac{2}{27} f^{(p,n)}_{TG} C_{11} \; .
\end{equation}
For the Majorana case of $O^M_{11}$, multiply the cross section from 
$O^D_{11}$ by a factor of 4.

For  $O^C_{15}$, the cross section is the same as $O^D_1$ with the 
following replacements 
\begin{eqnarray}
C^{u,d}_{1} &\longrightarrow & C^{u,d}_{15}  \; , \\
\Lambda_1 & \longrightarrow &\Lambda_{15} \; ,
\end{eqnarray}
for the couplings in Eqs.(\ref{sicouplings1a}) and (\ref{sicouplings1b}).
For  $O^C_{17}$, the cross section is same as $O^D_{7}$
with the following replacement 
\begin{eqnarray}
C^{u,d}_{7} &\longrightarrow & C^{u,d}_{17} \; , \\
\Lambda_7 & \longrightarrow &\Lambda_{17} \; ,
\end{eqnarray}
for the coupling in Eq.(\ref{sicouplings7}).
For  $O^C_{19}$,  the cross section is same as $O^D_{11}$ 
with the following replacement
\begin{eqnarray}
C_{11} &\longrightarrow & C_{19}  \; , \\
\Lambda_{11} & \longrightarrow &\Lambda_{19} \; ,
\end{eqnarray}
in  Eq.(\ref{fpnO11}).
The results for $O^R_{17,19}$ can by obtained by multiplying
a factor of 4 to the corresponding cross sections from $O^C_{17,19}$,
respectively.


\subsection{Spin-Dependent Cross Section}

For $O^D_4$, its contribution to the SD cross section is given 
by \cite{Engel:1992bf}
\begin{equation}
\label{sdxs4dirac}
\sigma^{\rm SD}_{\chi \mathcal N} (0) = \frac{8 \mu^2_{\chi \mathcal N}}{\pi}
G^2_F {\bar\Lambda}^2 J (J+1) \; ,
\end{equation}
where $J$ is the total spin of the nuclei $\mathcal N$, $G_F$ is the 
Fermi constant and
\begin{equation}
{\bar \Lambda} = \frac{1}{J} 
\left( 
a_p \langle S_p \rangle + a_n \langle S_n \rangle
\right) \; ,
\end{equation}
with $\langle S_p \rangle$ and $\langle S_n \rangle$ 
the average of the proton and neutron spins inside the
nuclei respectively, and
\begin{equation}
\label{sdcouplings}
a_{p,n} = \sum_{q=u,d,s} \frac{1}{\sqrt 2 G_F} \frac{C_4^q}{\Lambda_4^2} \Delta q^{(p,n)} \; ,
\end{equation}
with $\Delta q^{(p,n)}$ being the fraction of the spin carried by the quark $q$ 
inside the nucleon $p$ and $n$. For an updated analysis of $\Delta q^{(p,n)}$,
see Ref.\cite{Cheng-Chiang}.
For $O^M_4$, one should 
multiply the cross section Eq.(\ref{sdxs4dirac}) by
a factor of 4. 

For $O^D_5$, its cross section is 
the same as $O^D_4$ with the following replacements in Eq.(\ref{sdcouplings})
\begin{eqnarray}
C_4^q & \longrightarrow & 2 \, C_5^q \; , \\
\Lambda_4 & \longrightarrow & \Lambda_5 \; .
\end{eqnarray}
%


\section{Monojet and Monophoton production at Colliders}

Dark matter particles can be produced in hadronic collisions simply
by crossing the Feynman diagrams responsible for the SI or SD scattering 
between DM particles and nucleons. 
However, it would only give rise to something entirely
missing in the detection.  We therefore need some additional visible
particles for trigger.  One of the cleanest signatures is monojet or
monophoton production, which has only a high $p_T$ jet or photon
balanced by a large missing transverse momentum.  
The most precise measurements come from the CMS \cite{cms}
and the ATLAS \cite{atlas} experiments at the LHC.

In our approach of effective DM interactions, we can attach either a
gluon or a photon to one of the quark legs of the relevant operators.
For example, in $O_{1,7}$ we can attach a gluon or a
photon line to the fermion line.
For gluonic operators we can either attach a gluon line to the gluon leg
or attach the whole 4-point diagram to a quark line such that it
becomes a $q g$-initiated process.  The final state consists of a pair of DM
particles and a gluon or a photon. We require the jet or photon to
have a large transverse momentum according to the $p_T$ requirement of each
experiment.



For each effective operator $O_i$ we calculate the value of $\Lambda_i$
such that the SI cross section is about $1.9-2.0 \times 10^{-41} \,{\rm cm}^2$.
The results are shown in Table~\ref{SI-table}. Under the assumption that
the dark matter interacts universally with the quarks, the DM-nucleon
cross section is about the same for proton and neutron
(see Table~\ref{SI-table}). 
We use a dark matter mass $m_\chi =10$ GeV, and the results are not 
sensitive for  $m_\chi \sim 8-12$ GeV. 

\begin{table}[th!]
\caption{\small \label{SI-table}
The fitted values $\Lambda_i$ for the operators $O_{1,7,11,15,17,19}$, which
contribute to the spin-independent scattering between DM and nucleon.
The corresponding predictions for the number of monojet events for 
each operator
at LHC-8 for an integrated luminosity of 19.5 fb$^{-1}$ are also shown.
}
\begin{ruledtabular}
\begin{tabular}{cc|cc|cc}
Operators & $\Lambda _{i}$ &  \multicolumn{2}{c|}{$\sigma^{\rm SI}_{\chi N}$
                         ($\times 10^{-41}$ cm$^2$)} & 
    \multicolumn{2}{c}{ Number of Monojet events with $19.5$ fb$^{-1}$} \\ 
 &  (GeV)  & proton  &  neutron  & LHC-8  & Allowed/Ruled out \\
\hline
$O^D_1$ & 2500 &  $2.10$ & $2.11$  & $7.2$  & allowed   \\
$O_7^D$ & 85 & $2.00$  & $2.00$ &   $ 2.3$  & allowed  \\
$O^M_{7}$ & 106.4  & $ 2.12$ & $ 2.13$ & $ 1.3 $ & allowed \\
$O^D_{11}$ & 50.7 & $ 1.88$ & $ 1.88$ & $8.6\times 10^5$ & ruled out\\ 
$O^M_{11}$ & 63.8 & $1.88$ & $1.88$ &  $4.4\times10^5 $ & ruled out\\
$O^C_{15}$ & 2500 & $2.10$ & $2.11$ & $ 1.7 $ & allowed \\
$O^C_{17}$ & 175 & $2.00$ & $2.01$ & $ 1.8\times 10^{-3} $ & allowed
      \\
$O^R_{17}$ & 250 & $1.84$ & $ 1.88$ & $ 8.7\times 10^{-4} $ & allowed
 \\
$O^C_{19}$ & 117 & $ 1.89$ & $ 1.90$ & $332$ & allowed \\
$O^R_{19}$ & 147.3 & $ 1.89$ & $ 1.90$ & $166$ & allowed 
\end{tabular}
\end{ruledtabular}
\end{table}

The most recent monojet search was performed by the CMS collaboration
\cite{cms} with an integrated luminosity of 19.5 fb$^{-1}$.  It is almost
the full data set before the shutdown. The search for monojet events was
using the following selection cuts:
\begin{equation}
p_{T_{j}} > 110\;{\rm GeV},\;\; |\eta_j| < 2.4,\;\;
\not\!E_T > 250-550\;{\rm GeV} \;,
\end{equation} 
among which the 
\begin{equation}
\not\!E_T > 400\;{\rm GeV} \;
\end{equation}
was used specifically for the context of dark matter.  
In Ref.\cite{cms}, it was claimed that the
best expected limit was obtained with $\not\!E_T > 400\;{\rm GeV}$.
We therefore follow their claim and use 
$\not\!E_T > 400\;{\rm GeV}$. The observed upper limit on the 
number of events of the hypothetical signal of dark matter is
\begin{equation}
\label{434}
  N^{\rm obs} < 434 \;.
\end{equation}
We simply compare this observed upper limit of number of events to the
predictions implied by the CDMS II result. The numbers of monojet events 
for all SI operators are shown in the second last column of 
Table~\ref{SI-table}, while in
the last column we say ``allowed'' or ``ruled out'' 
as compared with Eq.(\ref{434}).

We note that our parton-level calculation gives similar numbers of events as
the dark matter model in the experimental paper \cite{cms}. 

\begin{table}[th!]
\caption{\small \label{SD-table}
The fitted values $\Lambda_i$ for the operators $O_{4,5}$, which
contribute to the spin-dependent scattering between DM and nucleon.
The corresponding predictions for numbers of monojet events
are also shown.
}
\begin{ruledtabular}
\begin{tabular}{cc|c|cc}
Operators & $\Lambda _{i}$ 
   & $\sigma^{\rm SD}_{\chi n}$ (neutron) 
   & \multicolumn{2}{c}{Number of Monojet events with $19.5$ fb$^{-1}$} \\ 
   & (GeV) & ($\times 10^{-36}$ cm$^2$) & LHC-8 & Allowed/Ruled out \\
\hline
$O^D_4$ & 28 & 8.93 & $ 4.5\times 10^{8} $ & ruled out \\
$O^M_4$ & 39.6 & 8.93 & $ 2.3\times 10^{8} $  & ruled out \\
$O^D_5$ & 28 & 8.93 & $3.6\times 10^9$  & ruled out
\end{tabular}
\end{ruledtabular}
\end{table}

We repeat the whole exercise for the SD cross section. It was shown
in Ref.~\cite{dep} that a SD scattering between the DM and the neutron
can explain the data with a SD cross section of $10^{-35}\,{\rm cm}^2$,
which is six orders of magnitude above the SI one.
We obtain the fitted parameters $\Lambda_i$ for $O_4^{D,M}$ and $O_5^D$ in
Table~\ref{SD-table}. The corresponding predictions for 
the number of monojet events
at the LHC-8 for these three operators are also shown in 
the second last column. 
It turns out the predicted numbers for monojet events are way too 
large compared with the experimental
upper limit in Eq.(\ref{434}). All these SD operators are ruled out.

In principle, one can also make use the monophoton event rates to get
bound on the DM interactions. Nevertheless, the results obtained with
monophoton are not as good as monojet at this stage.

In Table~\ref{LHC-14}, the monojet and monophoton cross sections
for the SI operators allowed by the current LHC-8 data are predicted for LHC-14.
The selection cuts on the monophoton events at the LHC-14 are
\begin{equation}
p_{T_{\gamma}} > 125\;{\rm GeV},\;\; | \eta_{\gamma} | < 1.5 \; , 
{\not\!E_T > 125 \,{\rm GeV}} \;,
\end{equation}
while the selection cuts for the monojet events are 
the same as those  for LHC-8.

\begin{table}[th!]
\caption{\small \label{LHC-14}
Predicted monojet and monophoton cross sections for operators 
$O_{1,7,15,17,19}$ at the LHC-14,  which are still allowed by 
current data at the LHC-8. 
Here ``-'' means that the gluonic operators do not give rise
to monophoton events in the first approximation.
}
\begin{ruledtabular}
\begin{tabular}{cccc}
Operators & $\Lambda _{i}$ (GeV) & Monojet cross section (fb) 
& Monophoton cross section (fb)  \\
\hline
$O^D_1$ & 2500 &  $4.9$ & $0.43$ \\
$O_7^D$ & 85 &    $14.3$  & $2.25$\\
$O^M_{7}$ & 106.4  & $7.5$ & $1.17$ \\
$O^C_{15}$ & 2500 & $1.1$  & $ 0.096$ \\
$O^C_{17}$ & 175 & $1.2\times 10^{-3}$ & $ 3.85\times 10^{-4}$       \\
$O^R_{17}$ & 250 & $5.6\times 10^{-4}$ & $ 1.85\times 10^{-4}$     \\
$O^C_{19}$ & 117 & $186$ & - \\
$O^R_{19}$ & 147.3 & $92.7$ & -  
\end{tabular}
\end{ruledtabular}
\end{table}

\section{Conclusions}

If the recent observation of three events by the CDMS II experiment is
interpreted as a 8.6 GeV dark matter signal, it would give corresponding
monojet/monophoton signals at the LHC.  We employed the effective DM
interaction approach and calculated the parameter that can account for
the observed cross section of the CDMS II events.  We found that the
current LHC-8 monojet data has already ruled out the SD operators
$O^{D,M}_4$ and $O^D_5$ that can be used to interpret the recent three
events from CDMS II by SD scattering.  One of the SI operators,
$O_{11}^{D,M}$, is also ruled by the LHC-8 monojet data.  
However, one must take these results with caution since 
for those operators that were ruled out by the current LHC data,
their best fitted effective scales are all less than 100 GeV. 
For such low scale, 
using the effective dark matter interaction may not be reliable 
at the LHC \cite{eftvalidity-1, eftvalidity-2}.
Nevertheless, the surviving SI operators $O_{1,7,15,17,19}$ from the 
current LHC-8 data
can be further probed in the LHC-14 run using the monojet as well as
the monophoton events.

\section*{Acknowledgments}
This work was supported in parts by the National Science Council of
Taiwan under Grant Nos. 99-2112-M-007-005-MY3, 102-2112-M-007-015-MY3, and
101-2112-M-001-005-MY3 as well as the
WCU program through the KOSEF funded by the MEST (R31-2008-000-10057-0). 
TCY would like to thank the hospitality of NCTS.


\begin{thebibliography}{99}

\bibitem{wmap} 
  G.~Hinshaw {\it et al.}  [WMAP Collaboration],
  arXiv:1212.5226 [astro-ph.CO].

\bibitem{planck} 
  P.~A.~R.~Ade {\it et al.}  [Planck Collaboration],
  arXiv:1303.5076 [astro-ph.CO].
  
\bibitem{dama}
R.~Bernabei {\it et al.}  [DAMA and LIBRA Collaborations],
  Eur.\ Phys.\ J.\ C {\bf 67}, 39 (2010)
  [arXiv:1002.1028 [astro-ph.GA]].

\bibitem{cogent}
C.~E.~Aalseth {\it et al.}  [CoGeNT Collaboration],
  Phys.\ Rev.\ Lett.\  {\bf 106}, 131301 (2011)
  [arXiv:1002.4703 [astro-ph.CO]].

\bibitem{cresst}
G.~Angloher,  M.~Bauer, I.~Bavykina, A.~Bento, C.~Bucci, C.~Ciemniak, G.~Deuter and F.~von Feilitzsch {\it et al.}
  Eur.\ Phys.\ J.\ C {\bf 72}, 1971 (2012)
  [arXiv:1109.0702 [astro-ph.CO]].

\bibitem{cdms}
R.~Agnese {\it et al.}  [CDMS Collaboration],
  [arXiv:1304.4279 [hep-ex]].

\bibitem{global}
K.~Cheung, P.~-Y.~Tseng, Y.~-L.~S.~Tsai and T.~-C.~Yuan,
  JCAP {\bf 1205}, 001 (2012)
  [arXiv:1201.3402 [hep-ph]].

\bibitem{Kurylov:2003ra} 
  A.~Kurylov and M.~Kamionkowski,
  Phys.\ Rev.\ D {\bf 69}, 063503 (2004)
  [hep-ph/0307185].

\bibitem{cao}
Q.~-H.~Cao, C.~-R.~Chen, C.~S.~Li and H.~Zhang,
  JHEP {\bf 1108}, 018 (2011)
  [arXiv:0912.4511 [hep-ph]].

\bibitem{lhc-3}
 J.~Goodman, M.~Ibe, A.~Rajaraman, W.~Shepherd, T.~M.~P.~Tait and H.~-B.~Yu,
  Phys.\ Rev.\  {\bf D82}, 116010 (2010)
  [arXiv:1008.1783 [hep-ph]].

\bibitem{lhc-5}
A.~Rajaraman, W.~Shepherd, T.~M.~P.~Tait and A.~M.~Wijangco,
  Phys.\ Rev.\ D {\bf 84}, 095013 (2011)
  [arXiv:1108.1196 [hep-ph]].

\bibitem{lhc-4}
  P.~J.~Fox, R.~Harnik, J.~Kopp and Y.~Tsai,
  Phys.\ Rev.\ D {\bf 85}, 056011 (2012)
  [arXiv:1109.4398 [hep-ph]].

\bibitem{ding-liao}    
 Ran Ding and Yi Liao 
  JHEP {\bf 1204}, 054 (2012)
  [arXiv:1201.0506 [hep-ph]].
          
\bibitem{cms}
CMS Collaboration, 
CMSPASEXO-12-048.

\bibitem{atlas}
ATLAS Collaboration, 
ATLAS-CONF-2011-096.

\bibitem{dep}
M.~R.~Buckley and W.~H.~Lippincott,
  arXiv:1306.2349 [hep-ph].

\bibitem{eftvalidity-1}
G.~Busoni, A.~De Simone, E.~Morgante and A.~Riotto,
  arXiv:1307.2253 [hep-ph].
    
\bibitem{eftvalidity-2}
S.~Profumo, W.~Shepherd and T.~Tait,
  arXiv:1307.6277 [hep-ph].

\bibitem{xenon}
E.~Aprile {\it et al.}  [XENON100 Collaboration], 
Phys.\ Rev.\ Lett.\  {\bf 107}, 131302 (2011) 
[arXiv:1104.2549 [astro-ph.CO]].

\bibitem{xenon-sd}
E.~Aprile {\it et al.} [XENON100 Collaboration],
 Phys. Rev. Lett. {\bf 111}, 021301 (2013);  \\
J.~Angle {\it et al.}  [XENON10 Collaboration],
  Phys.\ Rev.\ Lett.\  {\bf 101}, 091301 (2008)
  [arXiv:0805.2939 [astro-ph]].

\bibitem{zeplin}
V.~N.~Lebedenko {\it et al.} [ZEPLIN-III Collaboration],
  Phys.\ Rev.\ Lett.\ \ {\bf 103}, 151302  (2009)
  [arXiv:0901.4348 [hep-ex]].

\bibitem{simple}
M.~Felizardo {\it et al.} [SIMPLE Collaboration], 
  Phys.\ Rev.\ Lett.\  {\bf 108}, 201302 (2012)
  [arXiv:1106.3014 [astro-ph.CO]].

\bibitem{unbearable}
M.~T.~Frandsen, F.~Kahlhoefer, C.~McCabe, S.~Sarkar and K.~Schmidt-Hoberg,
 JCAP {\bf 07}, 023 (2013)
  [arXiv:1304.6066 [hep-ph]].

\bibitem{Cheng-Chiang} 
  H.~-Y.~Cheng and C.~-W.~Chiang,
  JHEP {\bf 1207}, 009 (2012)
  [arXiv:1202.1292 [hep-ph]].
  
\bibitem{Engel:1992bf} 
  J.~Engel, S.~Pittel and P.~Vogel,
  Int.\ J.\ Mod.\ Phys.\ E {\bf 1}, 1 (1992).

\end{thebibliography}
\end{document}